# Machine Learning Guided 3D Image Recognition for Carbonate Pore and Mineral Volumes Determination

Omar Alfarisi, Aikifa Raza, Hongtao Zhang, Djamel Ozzane, Mohamed Sassi, and TieJun Zhang

## Abstract

Automated image processing algorithms can improve the quality, efficiency, and consistency of classifying the morphology of heterogeneous carbonate rock and deal with a massive amount of data and images seamlessly. Geoscientists and petroleum engineers face difficulties setting the optimum method for determining petrophysical properties from core plug images of optical thin sections, Micro-Computed Tomography (μCT), or Magnetic Resonance Imaging (MRI). Most of the successful work is from the homogeneous and clastic rocks focusing on 2D images with less focus on 3D. Currently, image analysis methods converge to three approaches: image processing, artificial intelligence, and combined image processing with artificial intelligence. Our work proposes two ways to determine the porosity from 3D μCT and MRI images: an image processing method with Image Resolution Optimized Gaussian Algorithm (IROGA); an advanced image recognition method enabled by Machine Learning Difference of Gaussian Random Forest (MLDGRF).

Meanwhile, we have built reference 3D micro models and collected images for calibration of the IROGA and MLDGRF methods. To evaluate the predictive capability of these calibrated approaches, we ran them on 3D μCT and MRI images of natural heterogeneous carbonate rock. We also measured the porosity and lithology of the carbonate rock using three and two industry-standard ways, respectively, as reference values. Notably, IROGA and MLDGRF have produced porosity results with an accuracy of 96.2% and 97.1% on the training set and 91.7% and 94.4% on blind test validation, respectively, in comparison with the three experimental measurements. We measured limestone and pyrite reference values using two methods, X-ray powder diffraction and grain density measurements. MLDGRF has produced lithology (limestone and pyrite) volume fractions with an accuracy of 97.7% in comparison to reference measurements.

## Introduction

It is challenging to have static and dynamic reservoir properties determination and integration from micro and nano CT images with machine learning aided image recognition (*1-4*). The current literature review has revealed the pressing challenges regarding performing 3D image processing instead of 2D (*5-7*). Dual-energy μCT scanning data is used to characterize the rock properties (*8-10*), and claimed reasonable matching between measured core plug helium porosity, calculated using Digital Rock Physics (DRP) methods. The only drawback is the lack of blind test; instead, the work appeared as a learning and exploring μCT abilities to deliver quality images for rock characterization. (*9*) Work suggested that a 2D image cannot predict the petrophysical properties in carbonate due to the limited representation of heterogeneous rock in 2D. Lately, a major oil operator (*11*) stated that no commercial vendor could measure the porosity and permeability.

Besides, carbonate is more complex. In work performed (*11*) to check four DRP vendors' capabilities, the findings suggest no vendor managed to correctly get Porosity, Permeability, and $P_c$ (Capillary Pressure) curves. The work also recommended that future DRP analysis should also report the uncertainties. Artificial Intelligence (AI), mainly machine learning, was used to characterize rock properties (*12*) by analyzing μCT images, where segmentation is a method for porosity calculation. A Convolutional Neural Network (CNN) was used to get the pore size distribution. The use of 2D images (*12*), reminding us of the thin-sections analysis and highlights the effect of vugs that acts as large channels or high connectivity between matrix porosity, where the reported analysis is made using open-source ImageJ software (*13, 14*) and Tensor Flow software (*15*). While 3D Image analysis (*16*) for μCT images followed an approach called the Maximal Ball algorithm, as topological extraction of the pore network is the central part. The primary importance (*16*) is the estimation of the connectivity between each pore, knowing that quantifying connectivity is essential to calculate rock dynamic properties in putting a way forward an estimator for capillary pressure and relative permeability. NMR is a preferable porosity logging measurement in the oil industry (*17*), as it is lithology independent. The NMR signal represents rock total porosity for a fully saturated core plug (like oil or water). While if water (or oil) saturated core contains gas, then the gas is invisible to NMR. So, in the case of $CO_2$ flooding, the NMR will act as a liquid saturation tool. In (*18*), work focused on NMR for saturation measurement, the relation between relaxation time and pore size (*19*) was an evident link. Other research (*20*) earlier used the same methodology.

Therefore, in this work, we focused on Magnetic Resonance Imaging (MRI) to determine porosity and μCT to determine both porosity and lithology. As discussed earlier, carbonate is complex to characterize using image analysis (*8*) for determining porosity and permeability, concluding that image processing suits clastic only due to low heterogeneity. We have realized that tackling the challenge of determining porosity and lithology in heterogeneous rock requires an interdisciplinary approach. Delivering the solution requires fluid dynamics, reservoir characterization, image processing, machine learning, and software coding. Our research deployed new analysis and validation approaches, including building reference 3DMM (3D Micro Model) with various micropore sizes and using them as image processing calibration jig. Besides, we developed a new image resolution enhancement algorithm for high-quality segmentation. In our paper, we determined porosity using mainly two methods. The first is standalone image processing, where image feature extraction was successful. The second is machine learning-assisted image recognition. These methods depend on the training data from experimental results. We define Machine Learning as the ability of a machine to perform the learned tasks, as good as the training set, on new data. Human information indexing appears to be multi-contextual, while machine learning is single or contextually limited. So far, the machine performs specific tasks more effortlessly than a human can perform. Humans perform particular tasks more comfortably than machines, including image recognition (*21*) which is easier for humans than for machines. The machine can perform repetitive, rule-based mathematical tasks but lacks creativity and intuition (*22, 23*). Stepwise; human experts input training data to the machine and select learning algorithms; the machine performs the learning to build and test a model, then use the model to perform prediction on new data set, also known as supervised machine learning (*24*). Below, we give a background on the methods investigated: (1) machine learning, image processing, and (2) combined machine learning and image processing.

## Preliminaries

Coupling the latest science, engineering, and technology advancement became a necessary tool to solve long-standing challenges of rock properties characterization from 3D images; μCT and MRI.

In this section, we provide a brief description of the coupling between machine learning and image processing, starting by introducing each concept individually and then the combination of both, as per the following:

### 1- Machine Learning

Machine learning capabilities are vital for quality and efficiency, as the machine can reproduce results by mimicking what domain expert has taught. We have followed four steps in deploying machine learning:

(1) Step 1: Human expert (Geoscientist), a Petrophysicist in our case, provided the data set to the machine. Our petrophysicist identified two chemical (mineral) properties (Limestone and Pyrite), two physical properties (pore and solid), and various image intensities (*25-28*) corresponding to limestone, pyrite, and pore. The data set contains:

  i. "Independent parameters data columns"; image identified properties with black color, dark gray intensity, light gray intensity, and white.

  ii. "dependent data column,"; containing the desired classes of limestone, pyrite, and pore, while Solid here represents an aggregation of both lithology types; limestone and pyrite.

(2) Step 2: Human Expert chooses an open-source Machine Learning Algorithm (*29*); in our case, we used the "Random Forest" algorithm (*30*), owing to its superior capability in performing classification.
(3) Step 3: The computer (machine) uses the Random Forest algorithm to learn from the data set and build the prediction model with all its governing equations (*31*). The machine develops the required formulas that enable solving the assigned problem.
(4) Step 4: The computer uses the input data set, 3D μCT and MRI image stack, and deploys the learned prediction model with its governing equations to predict output classes.

**2- Image Processing**

We define image processing as the task a human expert performs on the image (2D or 3D) that produces a new set of information or a new image version to provide more insight. The image processing definition stepwise (*32*) appears as per the following:

(1) Step 1: Input an image (2D, 3D) to a machine. In our research, we focused on 2D and 3D images only where 2D represents surface in gray or color. At the same time, 3D represents the volumetric view (*33-36*).
(2) Step 2: Domain expert runs suitable filtering, segmentation, and analysis algorithms (*37-41*) to achieve the image-processing objectives.
(3) Step 3: The human expert learns new insight and generates new data (numerical or image) from the above step.

**3- Combining Machine Learning and Image Processing**

Machine Learning Image Recognition (MLIR) is defined as "the ability of a machine to perform tasks" (*42*) as good as the training set, image interpretation, delivered by a human expert, on new images faster and more accurate than a human expert. Another definition also describes machine learning and image processing (*43*). Image recognition machine learning (*44, 45*) requires a longer processing time in comparison with the processing time needed in performing image processing alone.

## Problem Articulation

Proper image classification (segmentation) of variant rock sections enables the quantification of its properties. However, segmentation accuracy depends heavily on how representative the image is to a real object. The pore sizes of our rock are either at or below the resolution of the image voxel size. In the following sections, we describe two challenges that we faced with 3D μCT and MRI images:

**1- First Challenge: Resolution Effect on 3D μCT Images**

Image resolution (*46*) is one of the main challenges in the 3D morphological (*47*) recognition process. Naturally, better resolution leads to better segmentation (*48, 49*). However, not every image comes with the desired resolution for optimal image analysis. In 3D rock images, identifying $10^6$ micro-pores and quantifying their morphology (*39, 50, 51*) are pressing challenges. The resolution does blur the feature of interest. Numerically, in blurring conditions, the pixel color value does not represent the actual color value for that position in space. We acquired 3D μCT images for dry (pore contains air only) carbonate rock and displayed one of its 2D slices in **Fig. 1**, where we see how micro-pores in carbonate rock appear in more than one gray value although it should appear as black. The cause of this challenge is low resolution, where the pore size is smaller than the pixel resolution used to acquire the image. The pore must be black, where the pixel should hold a value of "0". To accurately measure porosity and lithology (rock minerals), we must correct images to be "0" value for pore; otherwise, it is the rock-solid. The challenge is correcting without mistakenly converting a true-non-pore to a pore. In Fig. 1, the top image shows a 38 mm diameter sample acquired with 40 μm resolution (each pixel is 40 x 40 μm). The image inside the red circle is our zone of interest for identifying pore (Black) and solid (Light color or non-black). To solve the resolution effect, we used a convolutional Gaussian kernel for improving the resolution of 3D images.

## 2- Second Challenge: Resolution Effect on 3D MRI images

To identify the fluid distribution in the pore spaces, MRI can be one of the preferable options (*52, 53*). We used the NMR system of 0.5 Tesla. Besides, we proved, that MRI can also act as a porosity measuring tool independent of lithology type for saturated rock with single-phase fluid. The same carbonate rock sample was used to acquire 3D MRI images, as shown in Fig. **2**. The saturated pores appear a lighter color in MRI, while the solid appears dark (Black). We flooded the pores of a rock sample with crude oil. Fig. 2 shows three MRI cross-section images acquired with three different sampling rates and different slice thicknesses along the z-axis. Image voxel (x, y, z) resolution is (400 x 400 x 3500 μm) for Fig. 2a. While it is (400 x 400 x 2500 μm) for Fig. 2b image. And it is (400 x 400 x 1500 μm) for Fig. 2c. For porosity determination and fluid monitoring, it is important bringing Fig. 2c image (thinner 2D MRI slice, or higher resolution in the z-axis) to be clear as Fig. 2a image (thicker 2D slice or lower resolution in the z-axis). At the same time, we must improve the resolution of Fig. 2a to be as good as Fig. 2b. We notice that Fig. 2a has a higher intensity dynamic range despite its lower resolution (due to the thicker sampling size along the z-axis).

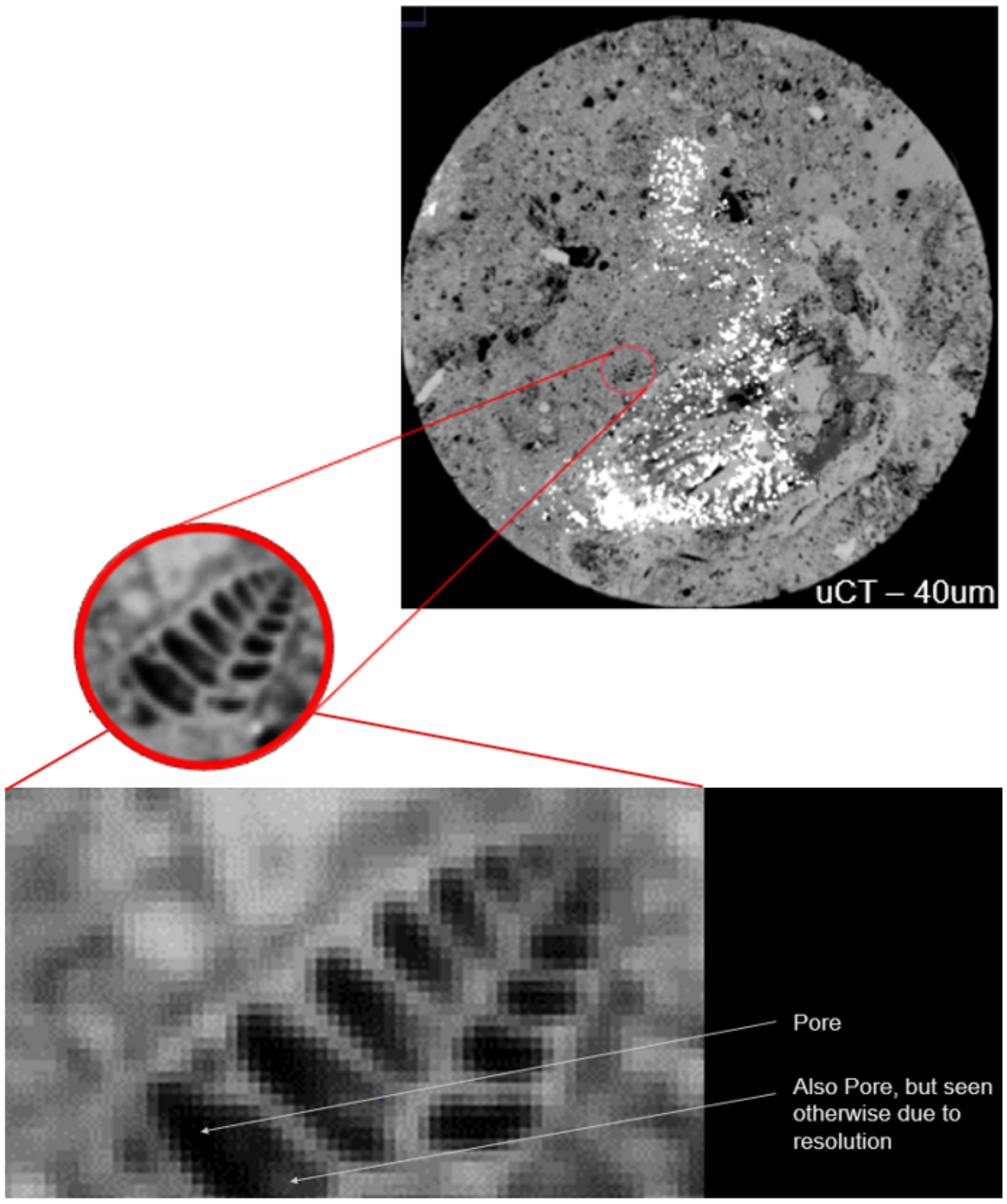

**Fig. 1–The resolution effect is evident in three different zooming scales.**

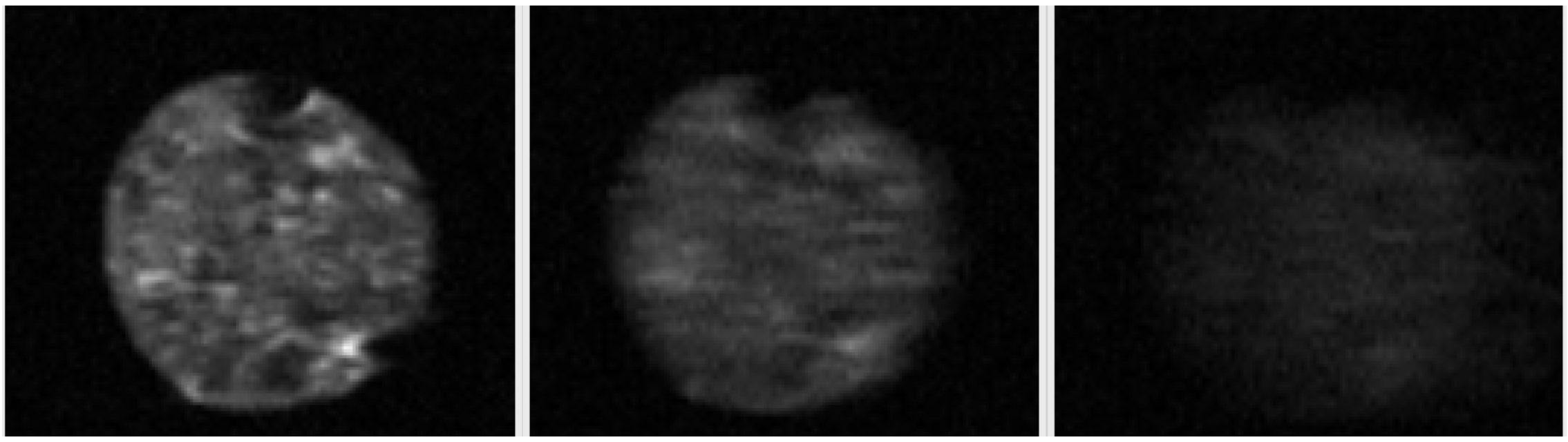

(a) (b) (c)

**Fig. 2–MRI (Magnetic Resonance Imaging) images for a carbonate core plug with three different sampling resolutions.**

## Proposed Methods

We focused on two methods for solving μCT and MRI images resolution) to analyze these images further to extract petrophysical properties. To achieve resolution enhancement and faster performance simultaneously, we chose image processing as a starting option. Then we compared the image processing results with what the machine learning approach can deliver. So, we conducted an experimental screening of two different methods. The sequence of our work in developing the best methodology for determining porosity and lithology is as described below:

### 1- First Method: Image Resolution Optimized Gaussian Algorithm (IROGA), an Image Processing Algorithm

A universal image enhancer provided an easy way for processing future MRI images without the need for human expert labeling; that is what we thought and then innovated and described in this paper. To address the resolution effects, we launched an experimental task that involves building a 3D Micro Model (3DMM). Using 3DMM MRI images as a calibration reference enabled us to produce universal image resolution enhancement, usable for similar acquisition devices, temperature, and acquisition parameters. Therefore, the same methodology we explain below is viable after conducting a new calibration run for a different device, environment, and acquisition parameters. We fabricated several micro models to quantify the blurriness of the image and calibrate it. In this paper, we briefly described one of the 3DMM bundles: the bundle of capillary tubes 3DMM. To increase the resolution by reducing the blurriness effect, we needed to build an image correction model, and at the same time, we had to ensure model performance efficiency. Therefore, a convolutional Gaussian image processing filter (*54, 55*) was the first choice to achieve better resolution. After several trial-and-error rounds, we developed a novel optimized convoluted Gaussian image processing algorithm, IROGA (Image Resolution Optimized Gaussian Algorithm). So, based on 3DMM MRI images, we constructed IROGA, as shown in **Fig. 3**. While segmenting (or interchangeably, we use "classifying") more than two classes, machine learning algorithms like the random forest (*4*) or others would be more suitable. We also used the words segments, labels, classes, clusters interchangeably as they mean the same in this paper. Let us keep in mind that IROGA as a standalone algorithm is useful when we target one or more of the following goals:

i. Enhance the resolution of an image.
ii. Differentiate two labels (binary).
iii. Perform points (i) and (ii) above together.
iv. Determine porosity from a low-resolution image.

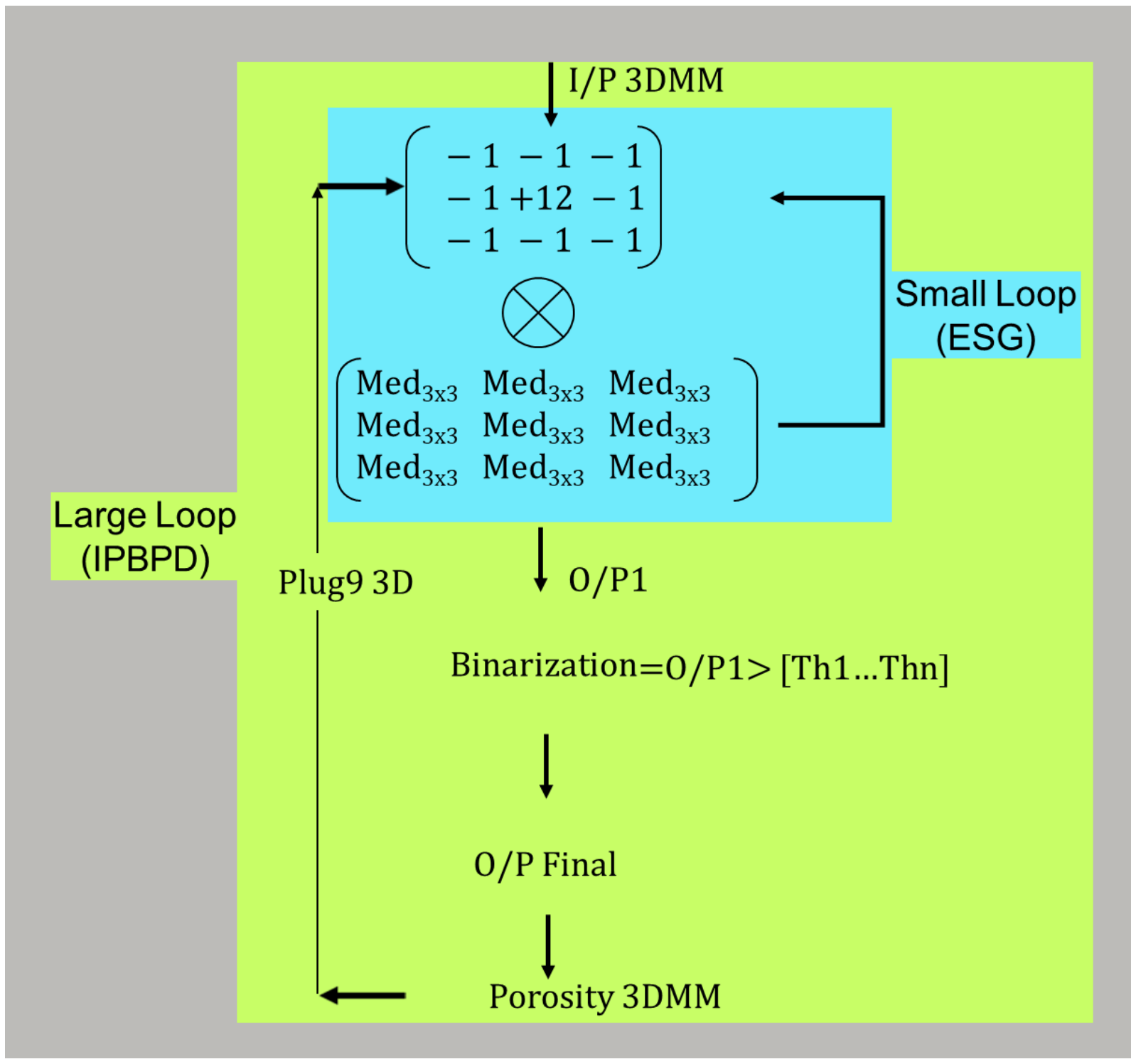

**Fig. 3–IROGA (Image Resolution Optimized Gaussian Algorithm) Block Diagram. Small Loop is also called Enhanced Convoluted Gaussian. Large Loop is also called Image Processing Based Porosity Determination.**

## 2- Second Method: Machine Learning Image Recognition (MLIR)

In the machine learning method, we performed 3D µCT image recognition using the Machine Learning Difference of Gaussian Random Forest (MLDGRF) algorithm. The primary machine learning performed tasks were as described earlier in the introduction section. The specific steps for porosity and lithology determination are that human experts provided pore, solid (limestone and pyrite) labels, and the Random Forest algorithm was selected to let the machine learn and then predict the segments. To do so, we used mainly three software; Microsoft Office 365™, Excel (*56*), Python (*57*), and ImageJ (*58*) for machine learning, image processing, mathematical modeling, and graph generation. **Fig. 4** shows the schematics of our proposed Machine Learning Image Recognition (MLIR) method. By describing the gradual composing of MLIR in three steps; identify optimal Machine Learning algorithms, deploy image recognition for targeted image feature, where image recognition is a branch of image processing (*43*). Then we combine both machine learning and image processing. **Fig.5** shows the MLIR algorithm description. Besides, to see the difference between Gaussian and Difference of Gaussian Function, we plotted both functions in **Fig. 6**.

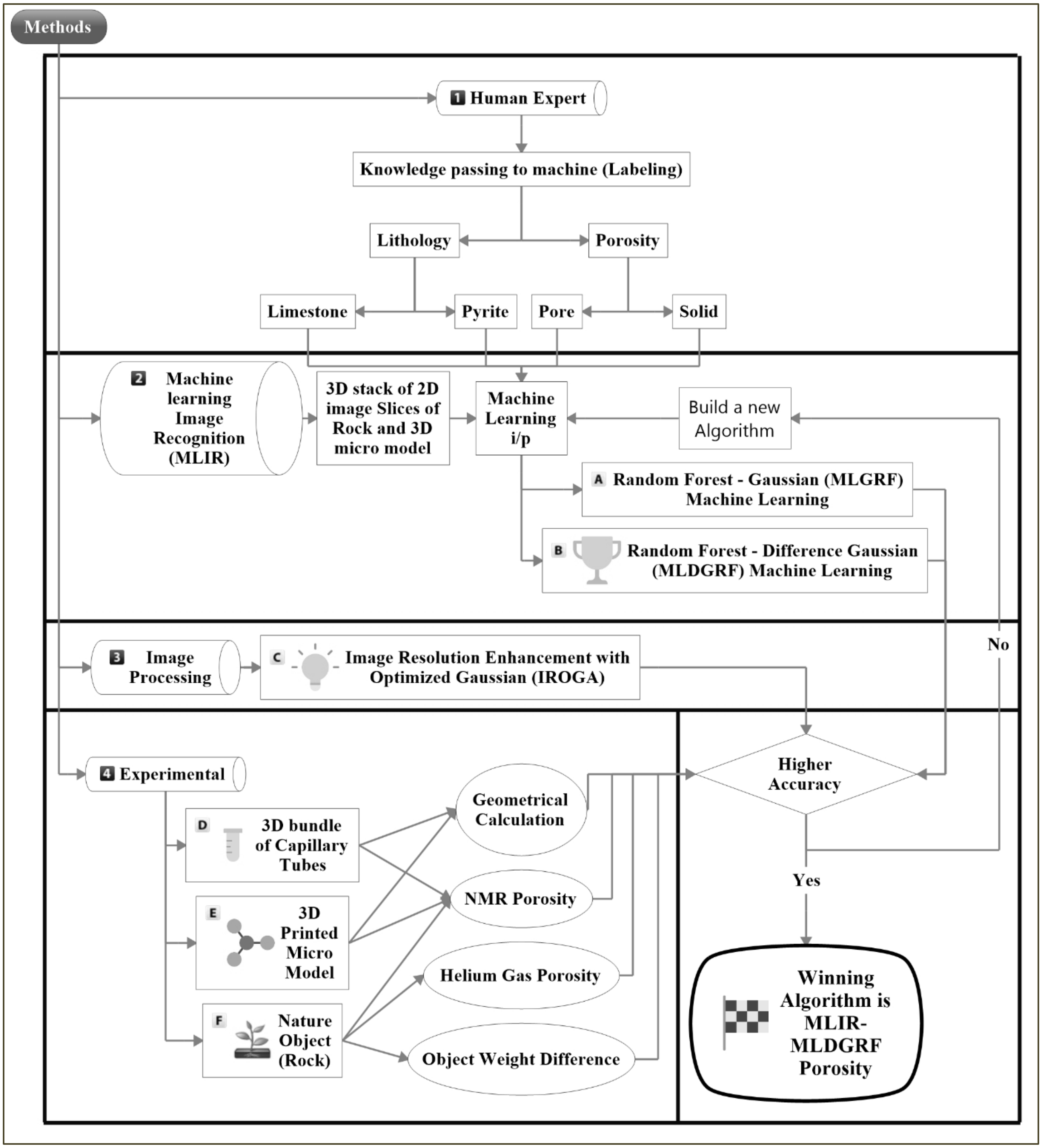

**Fig. 4–Algorithms of our 3D IROGA (Image Resolution Optimized Gaussian Algorithm) and MLIR (Machine Learning Image Recognition) porosity and lithology determination and validation.**

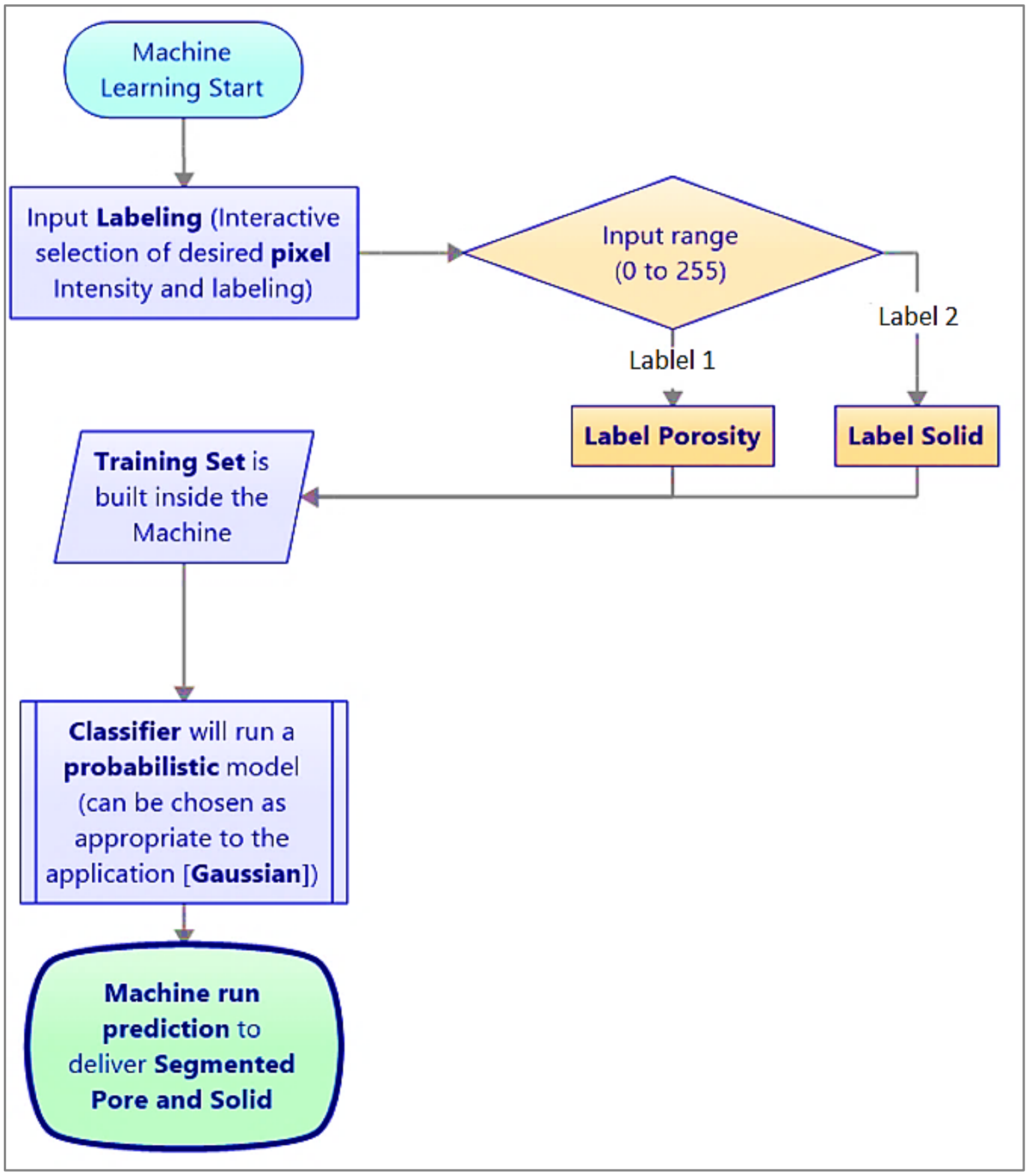

**Fig. 5–Machine Learning Flow Chart.**

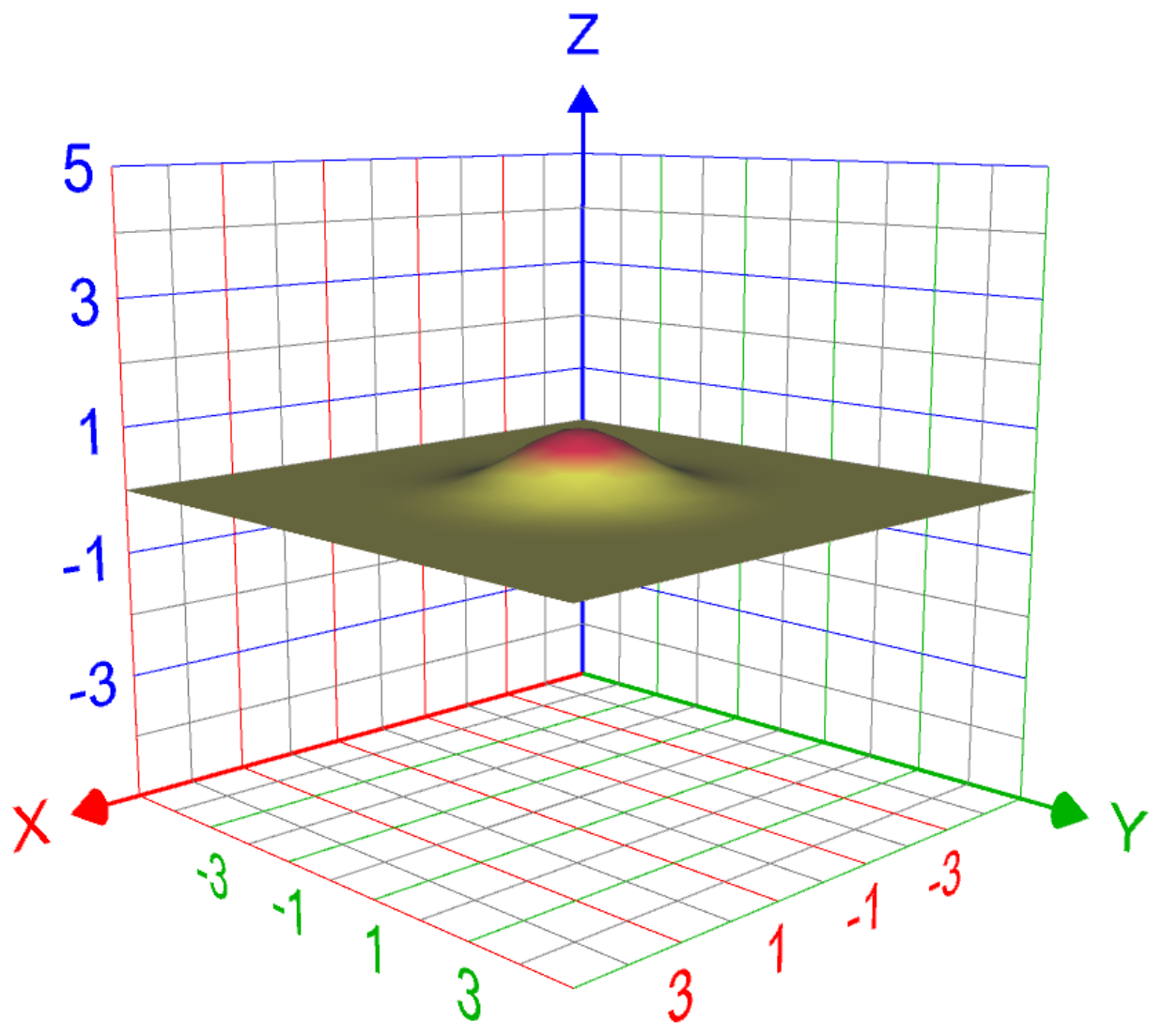

(a)

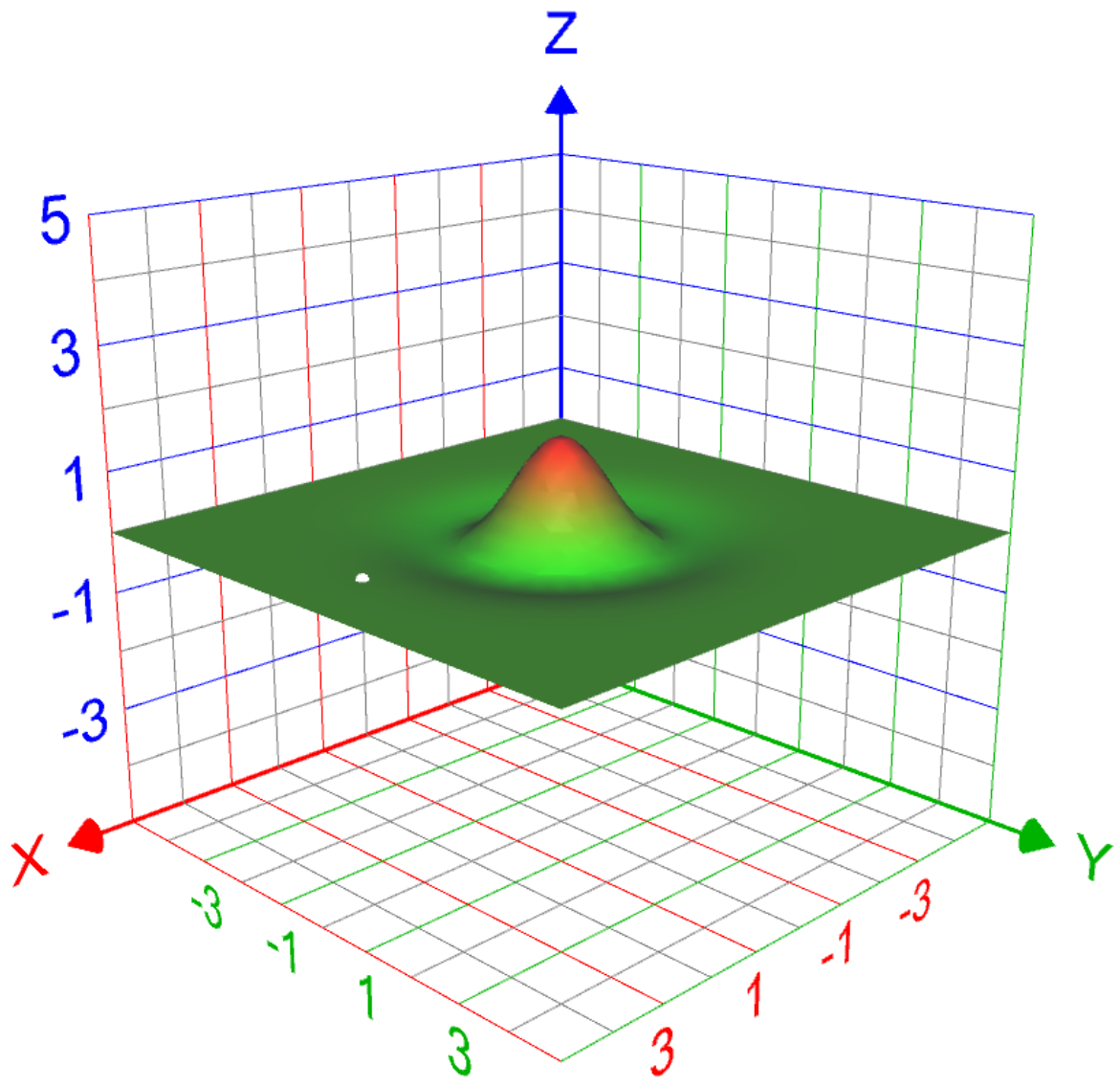

(b)

**Fig. 6–A 3D Graph of (a) Gaussian Function (b) Difference of Gaussian Function.**

## 3D Micro Model Experiments and Calibration

Without proper measurement control, the complexity of interpreting image data (μCT and MRI) will increase. Therefore, we developed 3D Micro Model (3DMM) with various micropore sizes to train and calibrate the proposed approaches. We went through several steps to generate 3DMM, measure, and calculate, as per the following:

### 1- 3DMM Construction

The first thing we started with was the development of the 3DMM. In this work, we use a bundle of capillary tubes 3DMM. We built 3DMM with known dimensions of pore and solid. It consists of 26 tubes, the length is 2 inches (50.8 mm), while the inner diameter of each tube is 600 μm. The 3DMM enabled constructing and validating IROGA in a controlled environment.

**Fig. 7** shows 3DMM with its cross-sectional and longitudinal structure. The geometrically calculated porosity of 3DMM is 7.21%.

### 2- 3DMM MRI Imaging and Porosity Calculations

We used MRI to image 3DMM, collecting a 3D MRI stack, which consists of nine 2D slices, as shown in **Fig. 8**. The small light dots, in Fig. 8, represent void (pore) filled with crude oil, each dot represents one tube while black represents solid objects, and the gray color rectangles represent the separators between each 2D image slice. We enlarged one of the nine MRI 2D slices, as displayed in **Fig. 9**.

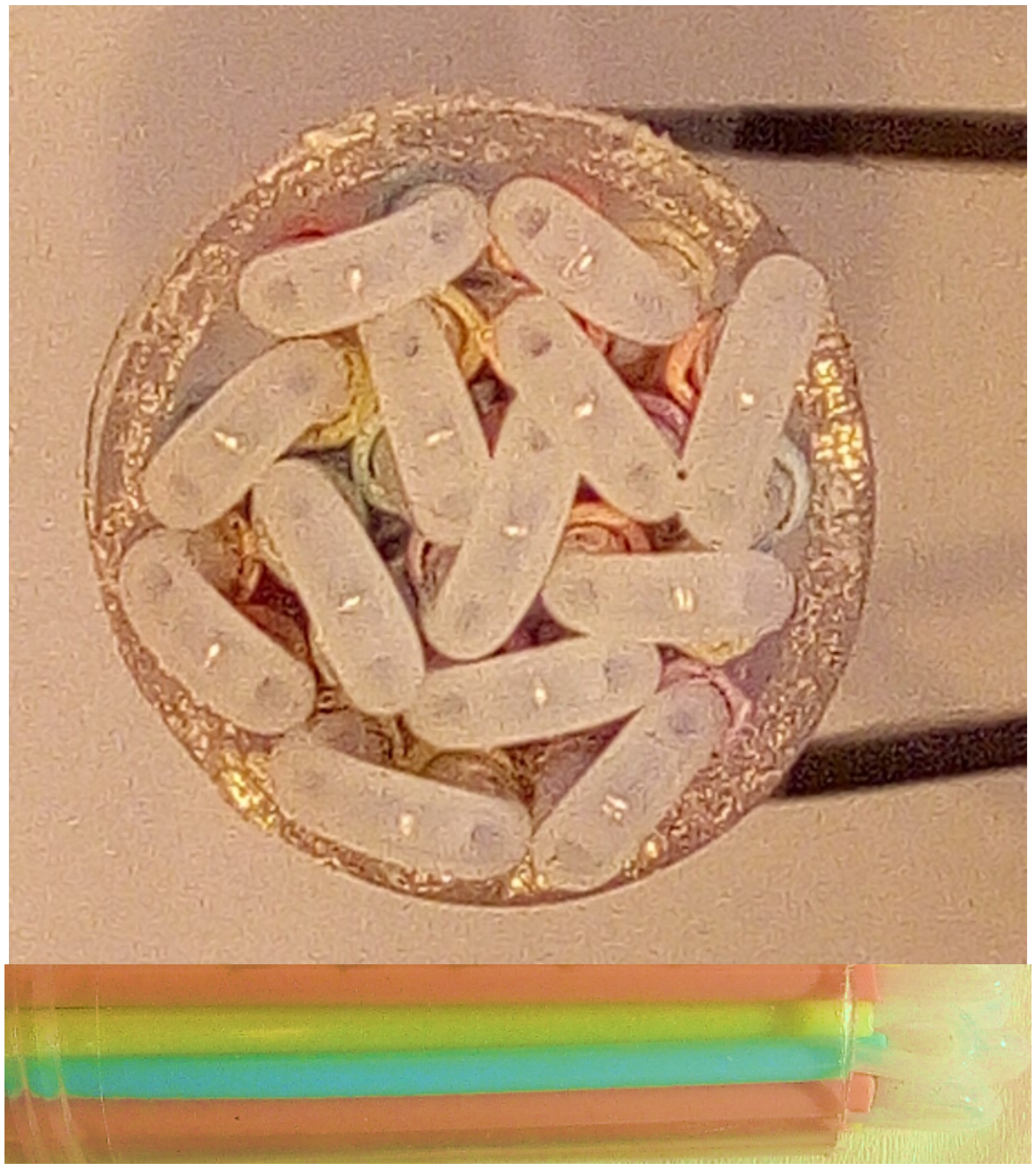

**Fig. 7–3D Micro Model (3DMM) with the (top) cross-section and (bottom) longitudinal section.**

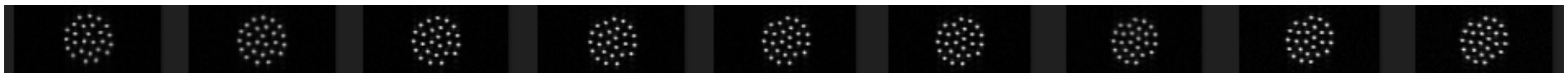

**Fig. 8–3D MRI cross-section image stack of nine 2D slices for the 3D Micro Model (3DMM).**

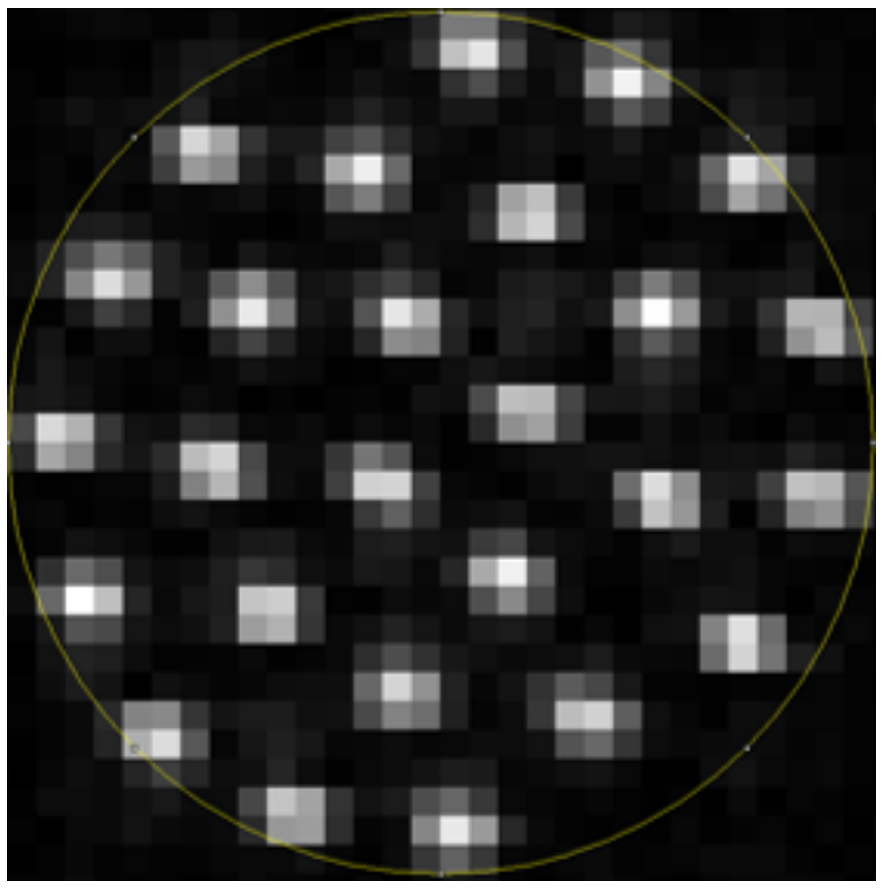

**Fig. 9–Enlarged MRI 2D slice of 3D Micro Model (3DMM) shows unwanted resolution defect in gray shade.**

The experiments can be used to train and calibrate our proposed methodologies as per below:

## 3- Calibration of Image Resolution Optimized Gaussian Algorithm (IROGA)

We designed IROGA after several verification steps. IROGA has two loops, as shown in Fig. 3, small loop, enhanced convoluted gaussian, and large loop, image processing porosity determination. We constructed IROGA with continuous iterations by optimizing the gaussian values of the small loop shown in Fig. 3. While **Fig. 10** is the final image of IROGA after applying the algorithm. We validated the IROGA model by calculating the porosity of 3DMM with an accuracy of 96.2%. We summarized below the development of IROGA modeling:

1- We built 3DMM with 600 µm inner diameter tubes (26 tubes).
2- We flooded 3DMM with crude oil collected from a lower Cretaceous Offshore Abu Dhabi formation.
3- We acquired 3D stack MRI Images with (400 x 400 x 3500 µm) resolution for the 3DMM.
4- We structured an iterative image resolution enhancement algorithm based on a convoluted Gaussian matrix.

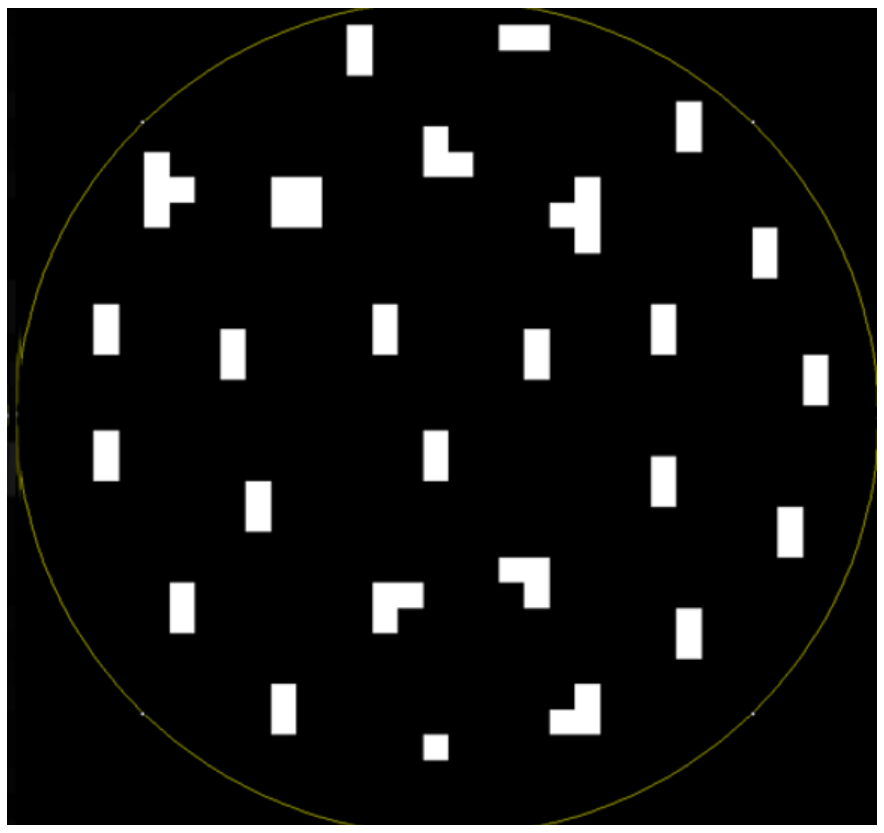

**Fig.10–Cross Section of one of the 2D slices of 3DMM MRI stack after applying IROGA (Image Resolution Optimized Gaussian Algorithm).**

## 4- Calibration of Machine Learning Difference of Gaussian Random Forest (MLDGRF)

We used the traditional Gaussian (*59*) Random Forest (*60*) as part of open-source ImageJ software (*14*). We trained the Gaussian algorithm with data set generated via interactive labeling using human expert knowledge. Then we selected the Gaussian function to build the best representative model and perform the classification. We noticed that the void (pore area) is more significant than our reference, and the accuracy is 53.1%, which is not acceptable. Therefore we used a different Random Forest algorithm, called Difference of Gaussian (*61-63*) Random Forest. In **Fig. 11**, we see the enhanced resolution image using

MLDGRF, where red represents solid, and the green represents pore. MLDGRF produced higher accuracy than IROGA to achieve 97.1%, the best accuracy method for resolution enhancement and porosity determination. Comparing MLDGRF with IROGA porosity results showed proper matching to reference porosity with 97.1% and 96.2% accuracy, respectively.

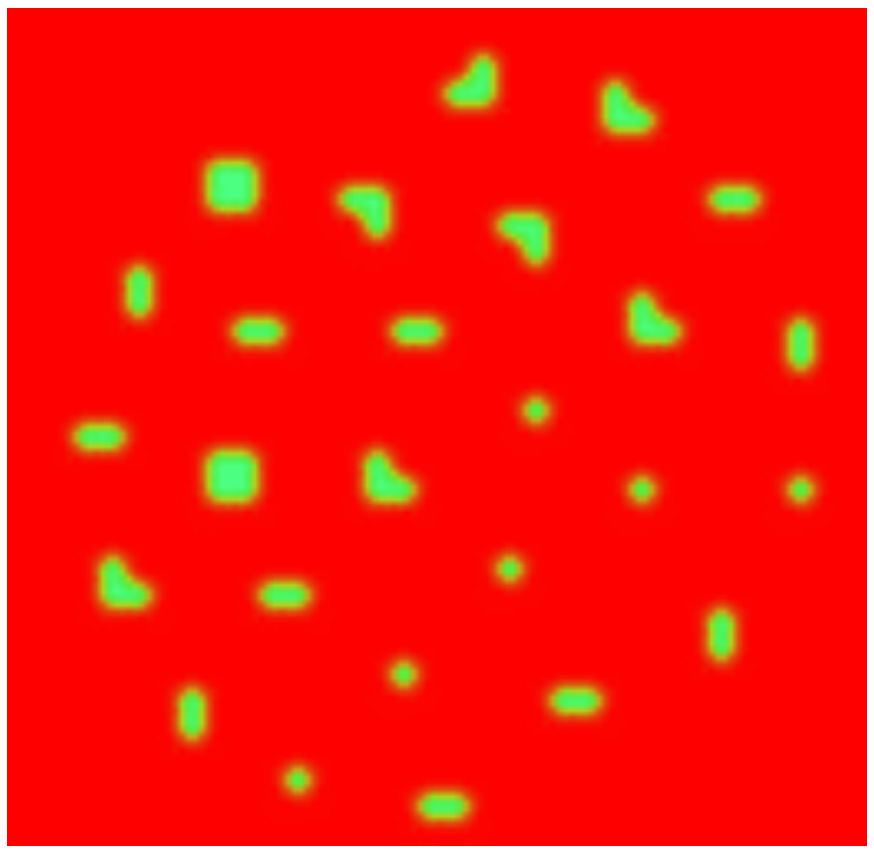

**Fig. 11–3DMM MRI image after using MLDGRF (Machine Learning Difference of Gaussian Random Forest), which we found to be the best methodology.**

## Results and Discussion

The goal of this work, as stated earlier, is to use the physical 3DMM calibrated models for determining the porosity and lithology of natural heterogeneous carbonate rock from 3D μCT and MRI images and using helium gas for core plugs (*64, 65*) of the natural heterogeneous carbonate rock. Also, we imaged the same core plug with μCT and MRI (*52*). We explain the steps of experimental porosity and lithology determination below:

### a) Natural Carbonate Rock Porosity: Experimental determination

We measured the core plug porosity with three different measurement methods. The first measurement method is rock's weight difference (*66*) under dry and wet conditions. We used NMR (*67*) on the same rock saturated with fluid (crude oil) in the second method. The third method is made with helium gas (*64, 65*), on the same rock sample after cleaning and drying, which is the ultimate reference of porosity value.

### b) Natural Carbonate Rock Lithology: Experimental Determination

Lab experiments define lithology (*68, 69*), where grain density measurement of rock core plug is essential to identify lithology. In addition to the grain density method, an X-Ray Powder Diffraction (XRPD) is another method to identify mineralogical fractions. The existence of pyrite in carbonate rock is visually recognizable to bare eyes due to its shiny metallic property. Limestone density is ~2.71 g/cm$^3$, while pyrite density is ~4.8 g/cm$^3$. The lab measurement showed the existence of two minerals using grain density calculation. At the same time, we also validated pyrite's existence by observing its appearance on the surface of the core plug. The higher the density of the mineral, the higher the voxel intensity of the μCT image. In our core plug, the pyrite density is much larger than that of limestone, producing identifiable shiny white regions on μCT images.

To ensure IROGA and MLDGRF method's capability to perform well with natural rock, we conducted validation tasks for both methods on lower Cretaceous heterogeneous carbonate rock. The validation results for both methods, IROGA and MLIR, are as per the following:

### 1- IROGA Porosity Validation with 3D MRI of Natural Carbonate Rock

We performed the validation of IROGA on MRI images of the carbonate core plug. **Fig. 12a** displays a raw MRI image of lower cretaceous carbonate rock saturated with crude oil. Black color represents solid, and light color represents pore filled with liquid.  Fig. 12b is the final MRI IROGA results in grayscale, and Fig. 12c is in color scale; the red color is solid, while the green color is pore filled with crude oil. For validating IROGA prediction, we used the reference helium gas porosity, and the resultant accuracy is 91.8%.

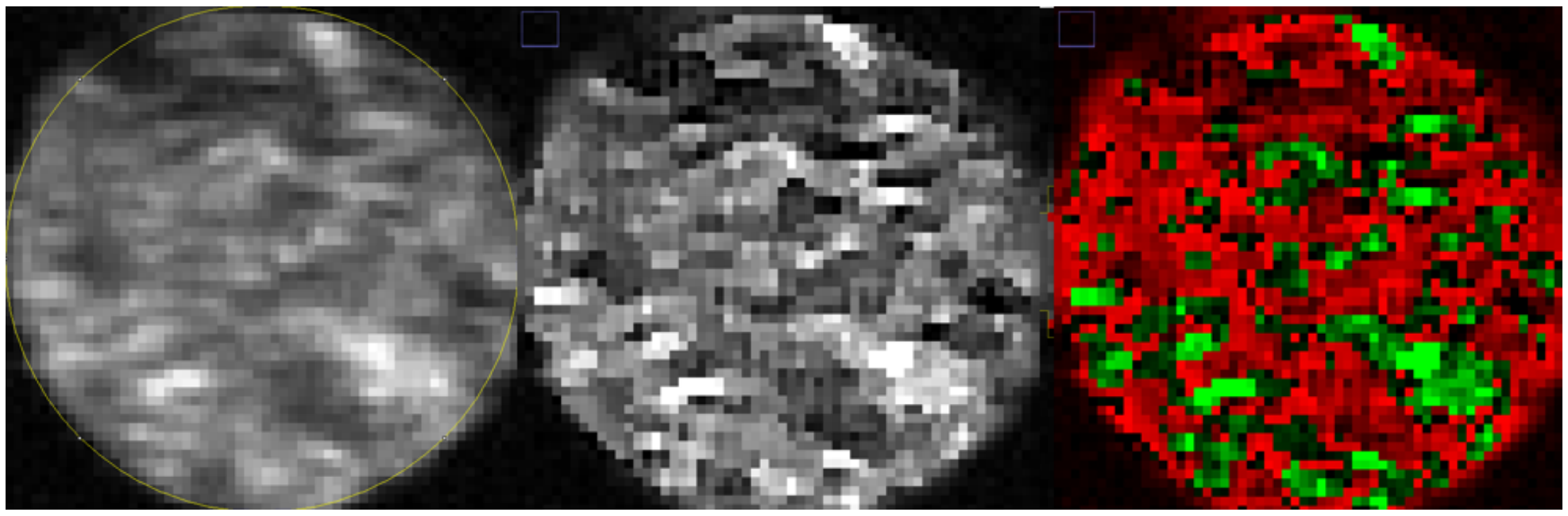

(a) (b) (c)

**Fig. 12–Cretaceous Carbonate Rock MRI; (a) Raw image, and (b) enhanced resolution image and (c) rock and oil separation.**

## 2- MLDGRF Porosity Validation with 3D µCT images of Natural Carbonate Rock

Since we have demonstrated the efficiency of MLIR, especially MLDGRF, in achieving superior accuracy when measuring 3DMM porosity, we extended the proposed approach to 3D µCT. We used the MLDGRF algorithm to predict 3D µCT porosity. We plotted the results of MLDGRF µCT Porosity in **Fig. 13a**. Then we compared MLDGRF results with three porosity measurements, as shown in Fig. 13b. The accuracy of MLDGRF reached 94.37%. The images of 3D µCT MLDGRF porosity prediction are in **Fig. 14**. The mismatch in porosity obtained from MLDGRF and the reference helium gas porosity is due to the performance of the Difference of Gaussian function parameters (*70, 71*). Therefore, we recommend further tuning the function parameters to increase the accuracy higher than 94.37%. Some low-level machine learning software (i.e., Python) (*72-74*) enable tuning of most of the parameters of Random Forest (*71*). At the same time, other higher-level machine learning software facilitates only a limited number of parameters to tune. We inferred that low-level machine learning software could help produce higher-quality results.

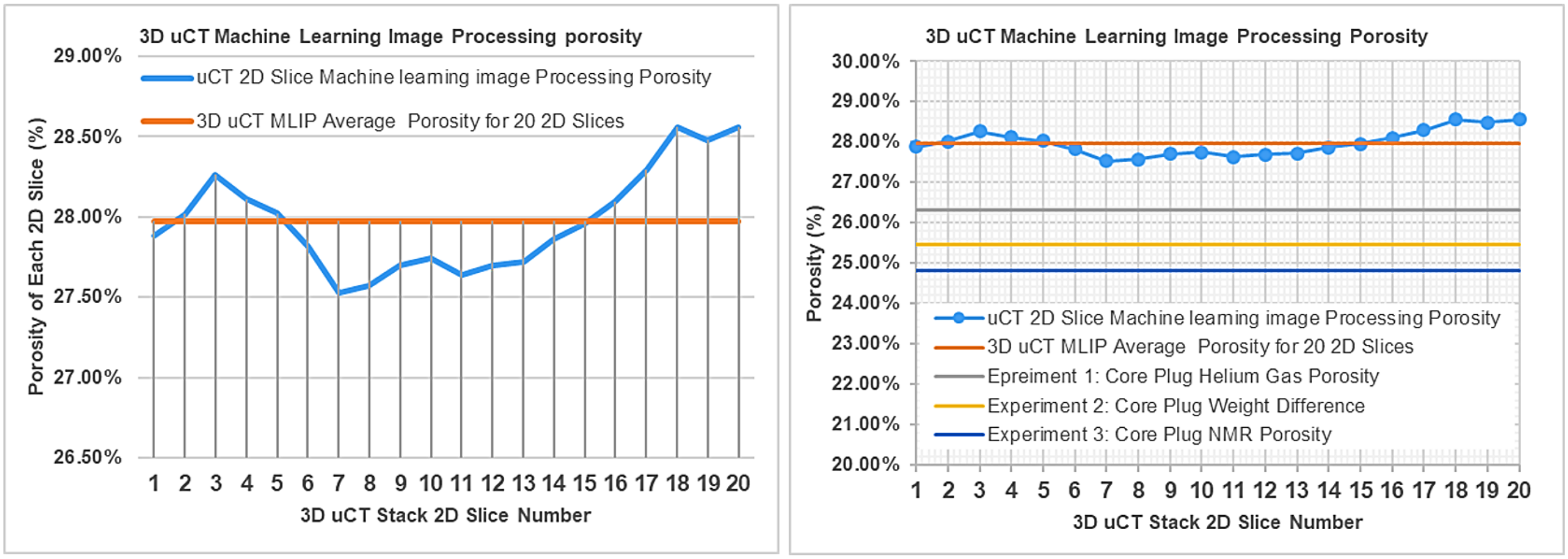

(a) (b)

**Fig. 13–Lower Cretaceous Carbonate Rock; (a) 3D µCT MLDGRF Porosity of 20 2D slices, (b) Comparing 3D µCT MLDGRF Porosity with reference porosities.**

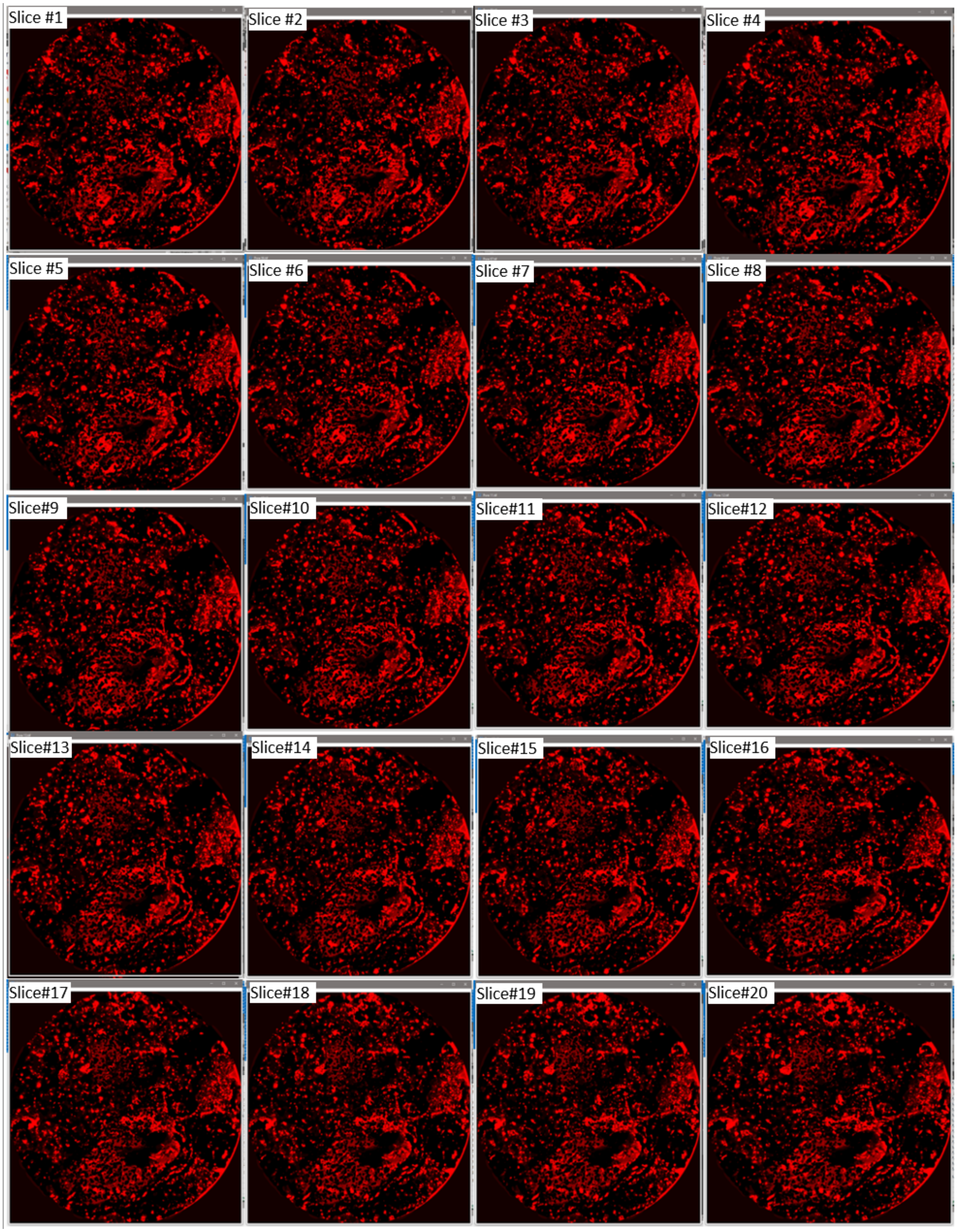

**Fig. 14–The Machine Learning Image Recognition predicted Porosity (red is pore) from 3D μCT carbonate rock image.**

## 3- MLDGRF Lithology Validation with 3D μCT Natural Heterogeneous Carbonate Rock

The vital step in performing rock typing as part of a reservoir characterization study is identifying lithology (*75-77*). In our research, human experts labeled different lithologies existing in 3D μCT stack, in our case, Pyrite and Limestone. Then the MLDGRF algorithm is applied to differentiate the minerals in the rock, see **Fig. 15**.

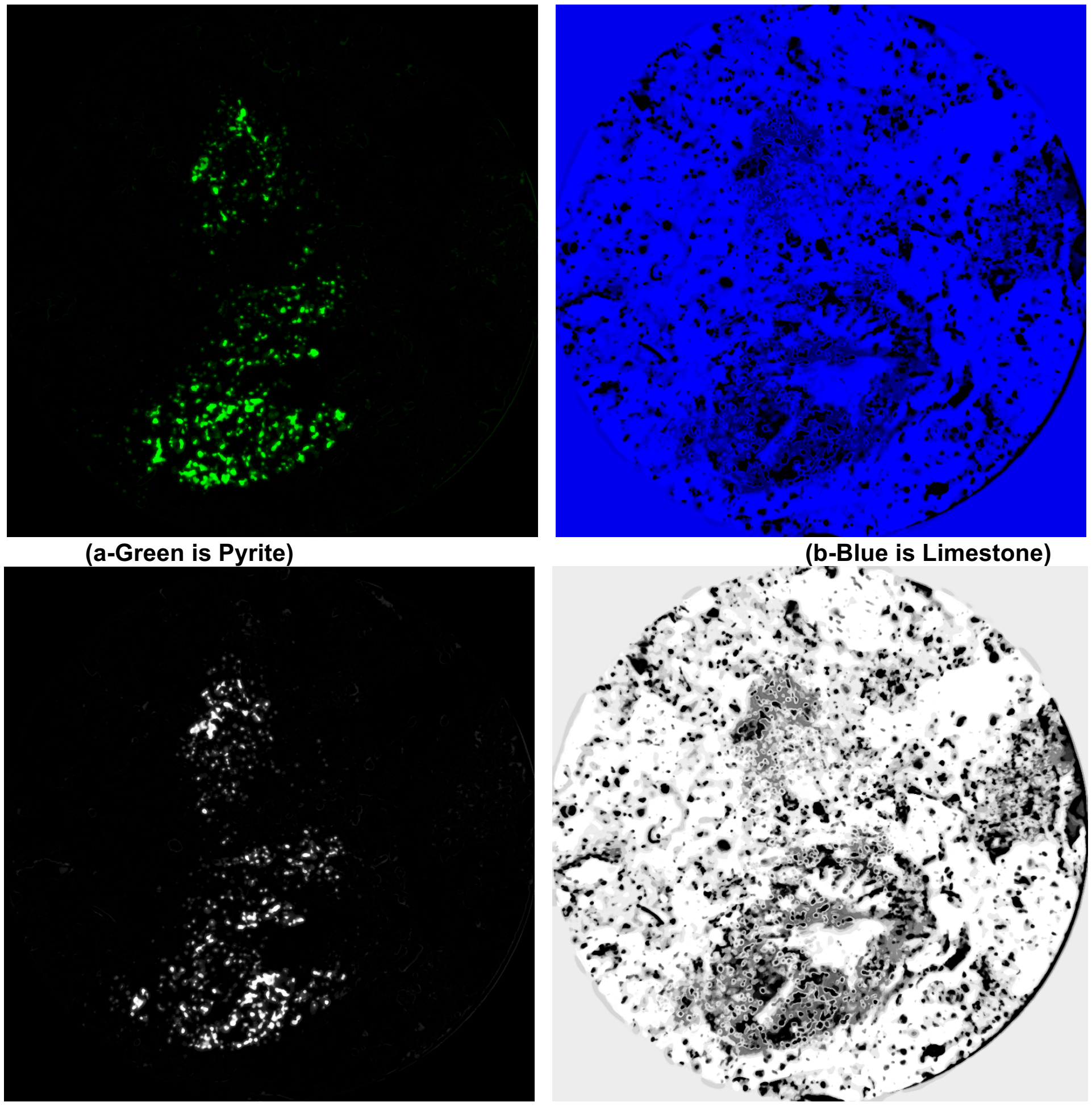

**Fig. 15–Machine Learning Image Recognition Difference Gaussian Random Forest algorithm MLDGRF has identified and segmented Lithology for 3D μCT Stack. Both (a-Green) and (a-White) are the same segmented image for Pyrite mineral, where Pyrite is the green color in (a-Green) image as the original segmentation image, and we displayed it again in (a-White) for printing purposes if the green color is not bright. While (b-Blue) and (b-White) are similar images for Limestone segmentation, where Limestone is the blue color in (b-Blue) image as the original segmentation image, and we displayed it again in (b-White) for printing purposes if the blue color is not bright enough.**

During the last 40 years, Geoscientists have performed mineral identification from images, even for outcrop rock (*78*). We noticed that most studies have worked on borehole imaging rather than a core plug μCT images (*79*). Currently, identifying and measuring minerals in core samples are mainly performed by physical weight measurements. While using imaging for lithology identification, either a thin section or SEM images provided the primary source for mineral analysis (*80*). This study focused on 3D μCT images to be the source of lithology identification and quantification. Most previous works focused on the textural identification for lithofacies and petrophysical properties (i.e., Archie's equation parameters and bioclast) (*79, 81-83*) determination. The main differentiator we used in our research for segmenting lithology was the image intensity. The image intensity of pyrite is ~255, which is a white color in a grayscale image 3D μCT and the highest intensity. There is no other mineral as dense as pyrite in our rock core plug, as the other mineral is limestone.

An external commercial laboratory performed the mineralogical analysis to reference rock mineral composition. The mineral lab measurement technique (*68*) uses helium gas for measuring void, outer volume geometry measurement, and the weight of the dry plug. This method cannot identify the location of pyrite inside the rock but can provide a value of grain density, which was ranging between 2.71 and 2.74 g/cm$^3$, a range that is higher than pure limestone. X-ray powder diffraction (XRPD) is another method, where the pyrite volume range is 0.1% to 1.1% out of total volume. At the same time, using 3D μCT, we can identify the existence and the distribution of mineral in the rock and a more accurate representation of volume. The XRPD method uses powder from one part of the rock, which does not represent the whole core plug. **Fig. 15** shows two different types of minerals (pyrite and limestone) that MLDGRF managed to segment and quantify. While **Fig. 16** shows the distribution of pyrite inside the rock in the 3D image caption, as identified by the MLDGRF. MLDGRF found a pyrite volume of 2.4% out of total volume, as shown in **Fig. 17**. MLRDGF produced an accuracy of 97.7% for pyrite determination. This identification of pyrite distribution helps permeability determination from 3D μCT and further rock classification.

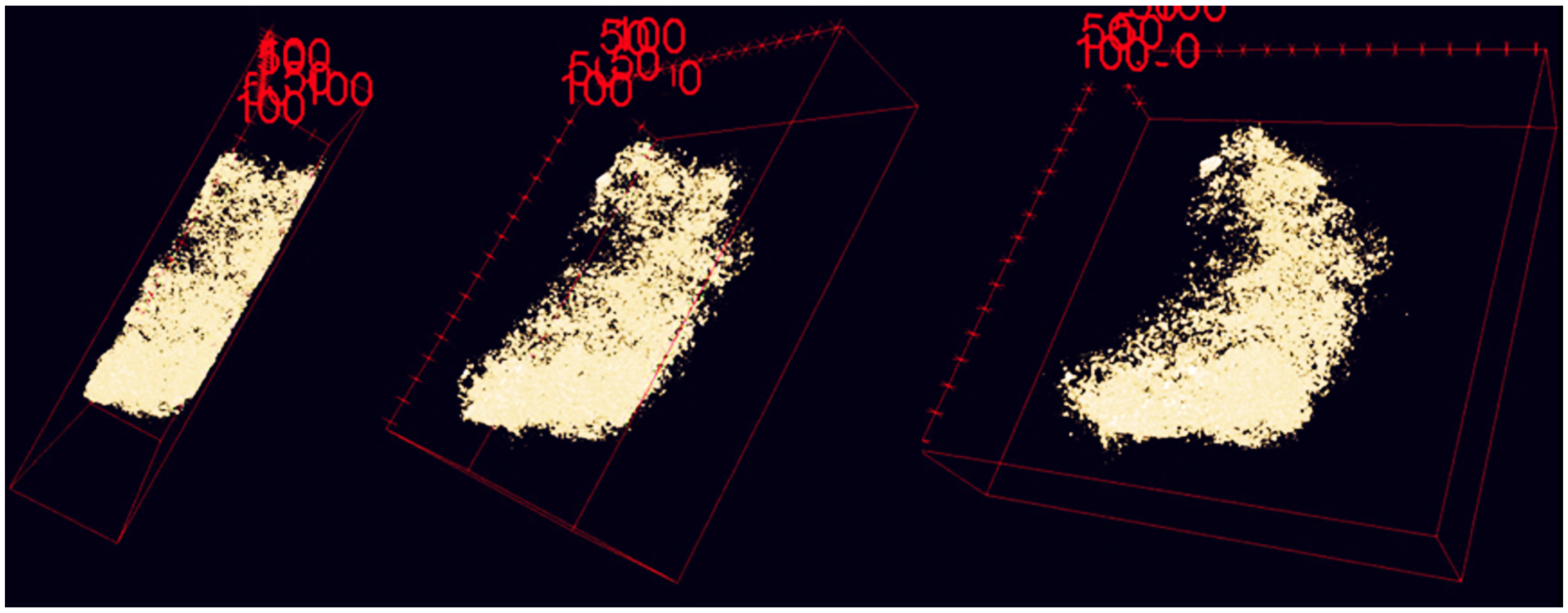

**Fig. 16–A 3D view caption of segmented pyrite using Machine Learning Image Recognition Difference Gaussian Random Forest algorithm MLDGRF for 3D μCT image Stack. The three 3D images show the same segmented pyrite from three different view angles.**

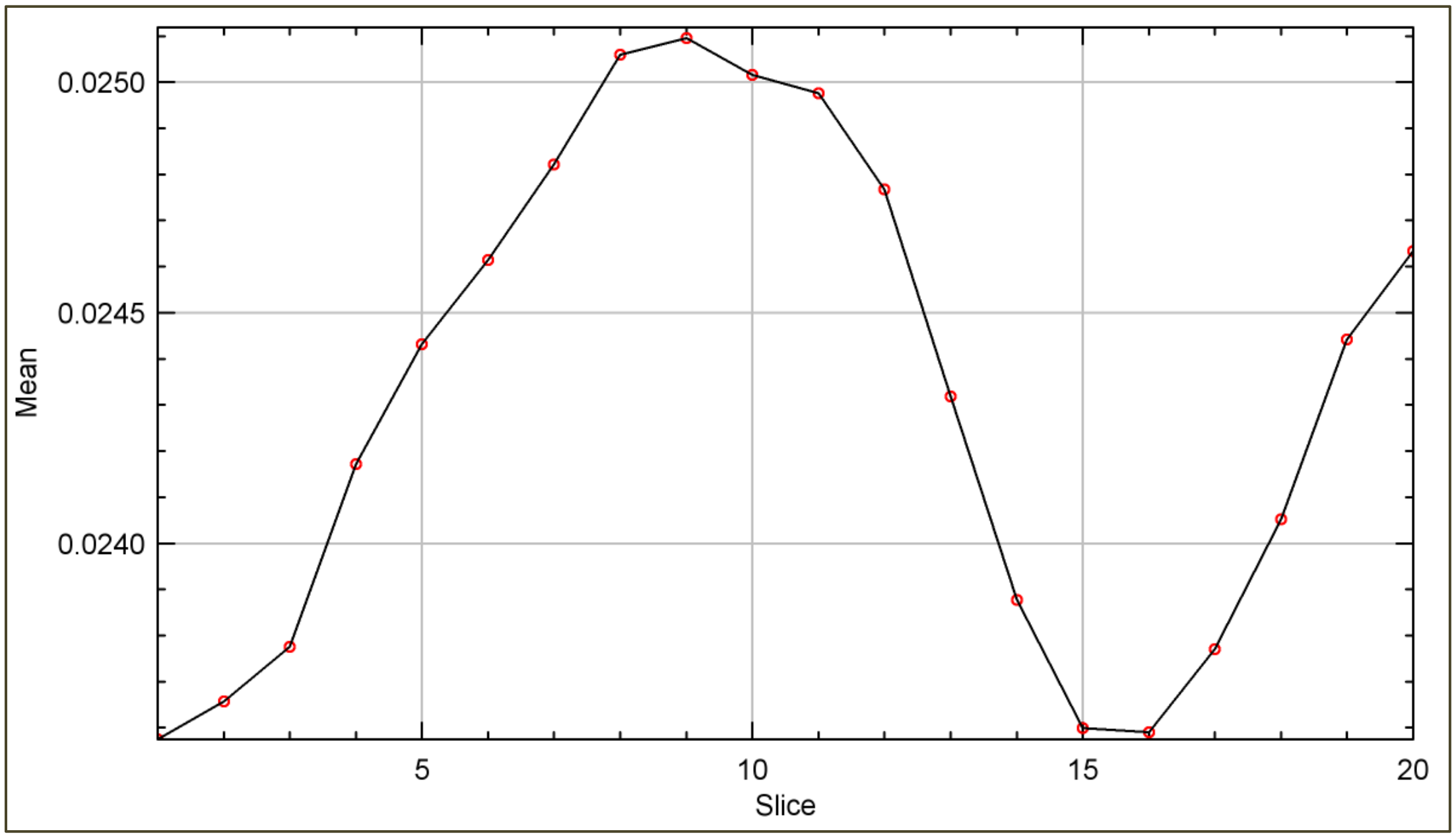

**Fig. 17– Lower Cretaceous Carbonate Rock 3D μCT MLDGRF Pyrite mean volume fraction for the 20 2D images slices.**

## Conclusions and Recommendations

1- To determine the reservoir properties (i.e., porosity and lithology), we investigated several methods and proposed a new one based on μCT and MRI images. We found that machine learning image recognition using the Difference of Gaussian Random Forest algorithm predicts better than machine learning image recognition using the Gaussian Random Forest algorithm. This finding helped us choose the optimal machine learning algorithm for the determination of reservoir rock from various image types. Moreover, experimental validation is necessary, as it is a physical proof and the reference to quantify the algorithm's accuracy.

2- Machine Learning Difference Gaussian Random Forest algorithm works well for different images (i.e., μCT and MRI). Machine learning image recognition can save years of tedious work if the domain expert correctly labels the desired features (labeling takes a few minutes) and chooses the best algorithm. Therefore, machine learning is recommended in more petroleum application scenarios to produce quality results faster.
3- Improving the resolution of 3D MRI images requires good image processing algorithms. We developed a new image enhancement algorithm for binary (two classes) segmentation with experimental work. Image processing can save time over machine learning for image recognition, where longer time is associated with a deeper machine learning algorithm. While for segmenting more than two classes, machine learning is better, especially with a heterogeneous rock.
4- Our work on porosity and lithology determination using machine learning has helped us better understand rock heterogeneity and provided insight into the analysis and digital classification of the reservoir rock.

## Affiliation

Omar Alfarisi - Khalifa University of Science and Technology & ADNOC Offshore.
Aikifa Raza - Khalifa University of Science and Technology.
Hongtao Zhang - Khalifa University of Science and Technology.
Djamel Ozzane - ADNOC.
Mohamed Sassi - Khalifa University of Science and Technology.
TieJun Zhang - Khalifa University of Science and Technology.

## Acknowledgment

We want to thank ADNOC, ADNOC Offshore, and Khalifa University for permitting the publication of this paper as part of a joint research project and a Ph.D. thesis. Especially we thank the support of Mr. Saoud Almehairbi, Ahmed Al-Hendi, Andreas Scheed, Ahmed Al-Riyami, Hamdan Al-Hammadi, Khalil Ibrahim, and Mohamed Abdulsalam. We shall also extend our acknowledgment to Prof. Isam Janajreh, Dr. Ashraf Al-Khatib, Dr. Zeyar Aung, Dr. Hongxia Li (Khalifa University of Science and Technology), Dr. Moustafa Dernaika (ADNOC), Mr. George Mani (Corelab), and Mr. Osama Jallad (Ingrain) for their support.

## References

1. T. G. Dietterich, in *International workshop on multiple classifier systems*. (Springer, 2000), pp. 1-15.
2. E. Decencière *et al.*, TeleOphta: Machine learning and image processing methods for teleophthalmology. *Irbm* **34**, 196-203 (2013).
3. I. Arganda-Carreras *et al.*, Trainable Weka Segmentation: a machine learning tool for microscopy pixel classification. *Bioinformatics* **33**, 2424-2426 (2017).
4. C. Sommer, C. Straehle, U. Koethe, F. A. Hamprecht, in *2011 IEEE international symposium on biomedical imaging: From nano to macro*. (IEEE, 2011), pp. 230-233.
5. C. I. Sánchez *et al.*, A novel automatic image processing algorithm for detection of hard exudates based on retinal image analysis. *Medical engineering & physics* **30**, 350-357 (2008).
6. H. Li, O. Chutatape, Automated feature extraction in color retinal images by a model based approach. *IEEE Transactions on biomedical engineering* **51**, 246-254 (2004).
7. A. C. Bovik, *The essential guide to image processing*. (Academic Press, 2009).
8. N. Saxena, G. Mavko, R. Hofmann, N. Srisutthiyakorn, Estimating permeability from thin sections

without reconstruction: Digital rock study of 3D properties from 2D images. *Computers & Geosciences* **102**, 79-99 (2017).

9. M. Dernaika *et al.*, Digital and Conventional Techniques to Study Permeability Heterogeneity in Complex Carbonate Rocks. *Petrophysics* **59**, 373-396 (2018).
10. G. C. Rafael, W. E. Richard, E. L. Steven, Digital image processing 3rd edition. *Eighth Impression*, (2007).
11. S. S. Chhatre *et al.*, in *Oral presentation given at the Annual Symposium of the Society of Core Analysts, Vienna, Austria*. (2017), vol. 27.
12. N. Alqahtani, R. T. Armstrong, P. Mostaghimi, in *SPE Asia Pacific oil and gas conference and exhibition*. (Society of Petroleum Engineers, 2018).
13. T. J. Collins, ImageJ for microscopy. *Biotechniques* **43**, S25-S30 (2007).
14. M. D. Abràmoff, P. J. Magalhães, S. J. Ram, Image processing with ImageJ. *Biophotonics international* **11**, 36-42 (2004).
15. M. Abadi *et al.*, in *12th {USENIX} Symposium on Operating Systems Design and Implementation ({OSDI} 16)*. (2016), pp. 265-283.
16. H. Dong, M. J. Blunt, Pore-network extraction from micro-computerized-tomography images. *Physical review E* **80**, 036307 (2009).
17. H. Xu, D. Tang, J. Zhao, S. Li, A precise measurement method for shale porosity with low-field nuclear magnetic resonance: A case study of the Carboniferous–Permian strata in the Linxing area, eastern Ordos Basin, China. *Fuel* **143**, 47-54 (2015).
18. Z.-Y. Liu, Y.-Q. Li, M.-H. Cui, F.-Y. Wang, A. Prasiddhianti, Pore-scale investigation of residual oil displacement in surfactant–polymer flooding using nuclear magnetic resonance experiments. *Petroleum Science* **13**, 91-99 (2016).
19. R. Ausbrooks, N. F. Hurley, A. May, D. G. Neese, in *SPE annual technical conference and exhibition*. (Society of Petroleum Engineers, 1999).
20. D. Marschall, J. Gardner, D. Mardon, G. Coates, in *1995 SCA Conference, paper*. (1995).
21. I. Goodfellow. (MIT Press, 2016).
22. L. Whitney, Are computers already smarter than humans? *Time Magazine, available at: http://time. com/4960778/computers-smarter-than-humans/(accessed 23 March 2018).[Google Scholar]*, (2017).
23. B. Marr. (2016, August 12).
24. S. B. Kotsiantis, I. Zaharakis, P. Pintelas, Supervised machine learning: A review of classification techniques. *Emerging artificial intelligence applications in computer engineering* **160**, 3-24 (2007).
25. E.-L. Chen, P.-C. Chung, C.-L. Chen, H.-M. Tsai, C.-I. Chang, An automatic diagnostic system for CT liver image classification. *IEEE transactions on biomedical engineering* **45**, 783-794 (1998).
26. G. Collewet, M. Strzelecki, F. Mariette, Influence of MRI acquisition protocols and image intensity normalization methods on texture classification. *Magnetic resonance imaging* **22**, 81-91 (2004).
27. D. W. Shattuck, S. R. Sandor-Leahy, K. A. Schaper, D. A. Rottenberg, R. M. Leahy, Magnetic resonance image tissue classification using a partial volume model. *NeuroImage* **13**, 856-876 (2001).
28. Y. Wang *et al.*, Computer-aided classification of mammographic masses using visually sensitive image features. *Journal of X-ray Science and technology* **25**, 171-186 (2017).
29. F. Pedregosa *et al.*, Scikit-learn: Machine learning in Python. *Journal of machine learning research* **12**, 2825-2830 (2011).
30. L. Breiman, Random forests. *Machine learning* **45**, 5-32 (2001).
31. R. Yu, Y. Yang, L. Yang, G. Han, O. Move, RAQ–a random forest approach for predicting air quality in urban sensing systems. *Sensors* **16**, 86 (2016).

32. J. C. Russ, *The image processing handbook*. (CRC press, 2016).
33. Y. Zhang, P. J. Passmore, R. H. Bayford, in *2005 IEEE Engineering in Medicine and Biology 27th Annual Conference*. (IEEE, 2006), pp. 1083-1086.
34. R. Deriche, Using Canny's criteria to derive a recursively implemented optimal edge detector. *International journal of computer vision* **1**, 167-187 (1987).
35. D. Dunlavy, T. G. Kolda, W. P. Kegelmeyer, "Tensor Decompositions for Analyzing Multi-link Graphs," (Sandia National Lab.(SNL-NM), Albuquerque, NM (United States); Sandia …, 2008).
36. J.-P. Calliess, M. Mai, S. Pfeiffer, On the computational benefit of tensor separation for high-dimensional discrete convolutions. *Multidimensional Systems and Signal Processing* **23**, 255-279 (2012).
37. A. Malcolm, H. Leong, A. Spowage, A. Shacklock, Image segmentation and analysis for porosity measurement. *Journal of materials Processing technology* **192**, 391-396 (2007).
38. W. B. Lindquist, S. M. Lee, D. A. Coker, K. W. Jones, P. Spanne, Medial axis analysis of void structure in three-dimensional tomographic images of porous media. *Journal of Geophysical Research: Solid Earth* **101**, 8297-8310 (1996).
39. R. Ehrlich, S. K. Kennedy, S. J. Crabtree, R. L. Cannon, Petrographic image analysis; I, Analysis of reservoir pore complexes. *Journal of Sedimentary Research* **54**, 1365-1378 (1984).
40. W. Lindquist, A. Venkatarangan, Investigating 3D geometry of porous media from high resolution images. *Physics and Chemistry of the Earth, Part A: Solid Earth and Geodesy* **24**, 593-599 (1999).
41. T. T. Mowers, D. A. Budd, Quantification of porosity and permeability reduction due to calcite cementation using computer-assisted petrographic image analysis techniques. *AAPG bulletin* **80**, 309-321 (1996).
42. B. H. Kann, R. Thompson, C. R. Thomas Jr, A. Dicker, S. Aneja, Artificial Intelligence in Oncology: Current Applications and Future Directions. *Oncology* **33**, (2019).
43. R. C. Gonzalez, R. E. Woods, S. L. Eddins. (Gatesmark Publishing, Knoxville, 2009).
44. K. He, X. Zhang, S. Ren, J. Sun, in *Proceedings of the IEEE conference on computer vision and pattern recognition*. (2016), pp. 770-778.
45. K. Simonyan, A. Zisserman, Very deep convolutional networks for large-scale image recognition. *arXiv preprint arXiv:1409.1556*, (2014).
46. J. Yang, J. Wright, T. S. Huang, Y. Ma, Image super-resolution via sparse representation. *IEEE transactions on image processing* **19**, 2861-2873 (2010).
47. B. Huang, W. Wang, M. Bates, X. Zhuang, Three-dimensional super-resolution imaging by stochastic optical reconstruction microscopy. *Science* **319**, 810-813 (2008).
48. W. Drexler *et al.*, Ultrahigh-resolution ophthalmic optical coherence tomography. *Nature medicine* **7**, 502 (2001).
49. J. Schindelin *et al.*, Fiji: an open-source platform for biological-image analysis. *Nature methods* **9**, 676 (2012).
50. K. M. Gerke, M. V. Karsanina, D. Mallants, Universal stochastic multiscale image fusion: an example application for shale rock. *Scientific reports* **5**, 15880 (2015).
51. K. Singh *et al.*, Dynamics of snap-off and pore-filling events during two-phase fluid flow in permeable media. *Scientific reports* **7**, 5192 (2017).
52. H. Wang *et al.*, Measurement and visualization of tight rock exposed to CO 2 using NMR relaxometry and MRI. *Scientific reports* **7**, 44354 (2017).
53. Q. Y. Zhou, J. Shimada, A. Sato, Three-dimensional spatial and temporal monitoring of soil water content using electrical resistivity tomography. *Water Resources Research* **37**, 273-285 (2001).
54. C. Li *et al.*, Analogue signal and image processing with large memristor crossbars. *Nature Electronics* **1**, 52 (2018).
55. A. Buades, B. Coll, J.-M. Morel, in *2005 IEEE Computer Society Conference on Computer Vision and Pattern Recognition (CVPR'05)*. (IEEE, 2005), vol. 2, pp. 60-65.
56. K. Wilson, in *Using office 365*. (Springer, 2014), pp. 1-14.

57. A. C. Müller, S. Guido, *Introduction to machine learning with Python: a guide for data scientists*. (" O'Reilly Media, Inc.", 2016).
58. T. Ferreira, W. Rasband, ImageJ user guide. *ImageJ/Fiji* **1**, 155-161 (2012).
59. C. E. Rasmussen, in *Summer School on Machine Learning*. (Springer, 2003), pp. 63-71.
60. D. F. Polan, S. L. Brady, R. A. Kaufman, Tissue segmentation of computed tomography images using a Random Forest algorithm: a feasibility study. *Physics in Medicine & Biology* **61**, 6553 (2016).
61. T. Lindeberg, Image matching using generalized scale-space interest points. *Journal of Mathematical Imaging and Vision* **52**, 3-36 (2015).
62. W. E. Polakowski *et al.*, Computer-aided breast cancer detection and diagnosis of masses using difference of Gaussians and derivative-based feature saliency. *IEEE transactions on medical imaging* **16**, 811-819 (1997).
63. D. G. Lowe, Distinctive image features from scale-invariant keypoints. *International journal of computer vision* **60**, 91-110 (2004).
64. S. C. Jones. (Google Patents, 1987).
65. S. C. Jones. (Google Patents, 1986).
66. D. Luffel, F. Guidry, New core analysis methods for measuring reservoir rock properties of Devonian shale. *Journal of Petroleum Technology* **44**, 1,184-181,190 (1992).
67. G. R. Coates, L. Xiao, M. G. Prammer, *NMR logging: principles and applications*. (Haliburton Energy Services Houston, 1999), vol. 234.
68. O. Torsæter, M. Abtahi, Experimental reservoir engineering laboratory workbook. *Department of Petroleum Engineering and Applied Geophysics, Norwegian University of Science and Technology (NTNU), Trondheim*, (2003).
69. H. Lee, S. Chough, Bulk density, void ratio, and porosity determined from average grain density and water content: an evaluation of errors. (1987).
70. R. Díaz-Uriarte, S. A. De Andres, Gene selection and classification of microarray data using random forest. *BMC bioinformatics* **7**, 3 (2006).
71. A. L. Boulesteix, S. Janitza, J. Kruppa, I. R. König, Overview of random forest methodology and practical guidance with emphasis on computational biology and bioinformatics. *Wiley Interdisciplinary Reviews: Data Mining and Knowledge Discovery* **2**, 493-507 (2012).
72. J. W. Peirce, PsychoPy—psychophysics software in Python. *Journal of neuroscience methods* **162**, 8-13 (2007).
73. M. F. Sanner, Python: a programming language for software integration and development. *J Mol Graph Model* **17**, 57-61 (1999).
74. T. E. Oliphant, Python for scientific computing. *Computing in Science & Engineering* **9**, 10-20 (2007).
75. O. Al-Farisi, A. Belgaied, H. Shebl, G. Al-Jefri, A. Barkawi, Well Logs: The Link Between Geology and Reservoir Performance. *Abstract Geo2002* **96**, (2002).
76. O. Al-Farisi, M. Elhami, A. Al-Felasi, F. Yammahi, S. Ghedan, in *SPE/EAGE Reservoir Characterization & Simulation Conference*. (2009).
77. O. Al-Farisi *et al.*, in *International Petroleum Technology Conference*. (International Petroleum Technology Conference, 2013).
78. T. H. Kurz *et al.*, Hyperspectral image analysis of different carbonate lithologies (limestone, karst and hydrothermal dolomites): the Pozalagua Quarry case study (Cantabria, North-west Spain). *Sedimentology* **59**, 623-645 (2012).
79. L. Knecht, B. Mathis, J.-P. Leduc, T. Vandenabeele, R. Di Cuia, in *SPWLA 44th annual logging symposium*. (Society of Petrophysicists and Well-Log Analysts, 2003).
80. J. He *et al.*, Quantitative microporosity evaluation using mercury injection and digital image analysis in tight carbonate rocks: A case study from the Ordovician in the Tazhong Palaeouplift, Tarim Basin, NW China. *Journal of Natural Gas Science and Engineering* **34**, 627-644 (2016).

81. R. A. Ketcham, G. J. Iturrino, Nondestructive high-resolution visualization and measurement of anisotropic effective porosity in complex lithologies using high-resolution X-ray computed tomography. *Journal of Hydrology* **302**, 92-106 (2005).
82. M. Knackstedt *et al.*, in *48th Annual Logging Symposium*. (Society of Petrophysicists and Well-Log Analysts, 2007).
83. M. Knackstedt *et al.*, in *SPE Asia Pacific Conference on Integrated Modelling for Asset Management*. (Society of Petroleum Engineers, 2004).